# Trajectory end point distribution of a test particle in the atmosphere

S.F Edwards  Cavendish Laboratory, Madingley Rd. CB3 0HE Cambridge, UK

Moshe Schwartz School of Physics and Astronomy, Sackler Faculty of Exact Sciences, Tel Aviv University, Ramat Aviv 69978, Israel

## Abstract

The classic meteorological law of diffusion in the atmosphere was given experimentally, by Richardson in 1926, whose result that the mean squared distance $\langle R^2 \rangle \propto T^3$, the time cubed, is in accord with the scaling theory of Komogorov [ Obukhov (1941)].
In some cases it might be important to have more information than that provided by Richardson's law. An example would be the distribution of pollutants in time by turbulent flow. Here small amounts of material reaching relatively large distances are of importance. This motivates our interest in the full distribution of the location of particles swept by the fluid as a function of time. The distribution depends on the distance through the dimensionless quantity $X^2 = \mathbf{R}^2 / \langle R^2(T) \rangle$. Using the Kolmogorov picture, we find that for small $X$, the distribution $P(\mathbf{R},T)$ is proportional to $\exp(-aX^2)$ and $\exp(-bX^{4/3})$ at its tail when $X$ is large.

A test particle swept around by turbulent air was shown experimentally by Richardson in 1926 [1] to satisfy the law that the mean squared displacement, $\langle R^2 \rangle$ after a time $T$ is proportional to the time cubed: $\langle R^2 \rangle \propto T^3$. This law was first derived by Obukhov [2] and is in accord with Kolmogorov's scaling theory of turbulence [3,4]. Later different derivations were given by Batchelor [5] and Lin [6]. The Richardson Law tells us something about the way a localized dust cloud is spread in time by turbulent air. If that dust cloud is a dangerous pollutant, for example, we would like to have much more detailed information than that given by the Richardson law. Such information is encoded in the probability distribution of the displacement $\mathbf{R}$ of the particle as a function of $T$, $P(\mathbf{R},T)$. The probability distribution, $P$, is not given by previous theoretical derivations [2,5,6], who only give

$$\langle R^2(T) \rangle = \int d\mathbf{R}\, \mathbf{R}^2 P(\mathbf{R},T) = aT^3, \qquad (1)$$

The purpose of the present article is to give the form of the displacement distribution in two extreme cases $\mathbf{R}^2/aT^3 \ll 1$ and $\mathbf{R}^2/aT^3 \gg 1$.

The equation describing the motion of the swept particle is given by

$$\dot{\mathbf{R}} = \mathbf{v}(\mathbf{R}(t),t), \qquad (2)$$

where $\mathbf{v}(\mathbf{r},t)$ is the given velocity field that carries the test particle. The actual velocity field is assumed to be random with zero mean and given correlations,

$$\langle \mathbf{v}(\mathbf{r_1},t_1) \cdot \mathbf{v}(\mathbf{r_2},t_2) \rangle = \phi(|\mathbf{r_1}-\mathbf{r_2}|,|t_1-t_2|) \equiv \phi(|\mathbf{r}_{12}|,t_{12}), \qquad (3)$$

We will proceed at present with the general form above and introduce the correlations specific to a turbulent fluid later.

We consider the space of trajectories connecting the origin at time $t=0$ and ending and some specified end point $\mathbf{R}$ at time $T$. Our first step is to find within that space the probability distribution of a particular trajectory, $\mathbf{r}(t)$ In doing so we will assume at present that the Kolmogorov theory [3,4] is exact and consequently so is Richardson's law. We are aware, of course, that modern theory and experiment yield results which slightly deviate from Kolmogorov [7-13] but our present aim is to produce a picture of the distribution function which is as simple as possible. We expect, however, to present the finer details in a subsequent publication.

We start by noting that the Richardson result can be obtained from a simplified theory in which the turbulence gives random impulses to the **velocity** of the particle in contrast to the Einstein random walk, which gives random jumps to the **position**. In terms of path integrals, the Einstein random walk, as described by Feynman has the trajectory distribution

$$P\{\mathbf{r}(t)\} \propto \exp[-\frac{1}{2D}\int_0^T \dot{\mathbf{r}}^2(t)dt], \qquad (4)$$

leading to

$$P(\mathbf{R},T) = (\frac{1}{2\pi DT})^{3/2} \exp(-\mathbf{R}^2/2DT) \qquad (5)$$

and hence

$$\langle R^2(t) \rangle = DT. \qquad (6)$$

The distribution coming from random jumps of the velocity is thus given by

$$P\{\mathbf{r}(t)\} \propto \exp[-\frac{1}{2Q}\int_0^T \ddot{\mathbf{r}}^2(t)dt], \qquad (7)$$

which gives the Richardson relation (1), $\langle R^2(T) \rangle = QT^3$ and

$$P(\mathbf{R},T) = (\frac{1}{2\pi QT^3})^{3/2} \exp(-\mathbf{R}^2/2QT^3) \ . \tag{8}$$

We will refine now the crude picture given above and obtain the trajectory end point distribution from the velocity correlations given by equation (3). We will assume that the functional velocity distribution is Gaussian, defined by the correlations given above,

$$\Gamma\{\mathbf{v}(\mathbf{r},t)\} \propto \exp[-\frac{1}{2}\int d\mathbf{r}_1 d\mathbf{r}_2 dt_1 dt_2 \mathbf{v}(\mathbf{r}_1,t_1)\cdot \phi^{-1}(|\mathbf{r}_{12}|,t_{12})\mathbf{v}(\mathbf{r},t_2)], \tag{9}$$

where $\phi^{-1}$ is the matrix inverse of the velocity correlations (3). This assumption seems adequate, although not exact, in two extreme cases, the case where the fluid is driven by noise which is extremely infrared in space or by noise which is white in space [14]. Experiments on real turbulent flow [7,9], which must be characterized by an additional length, corresponding to the inverse momentum scale, $\Lambda$, over which energy is pumped into the system, suggest, however, that the distribution function given above is an over simplification. Following Kraichnan, who used it in the passive scalar problem [15], we will also use, however, the above Gaussian distribution. The effect on the trajectory end point distribution due to deviations from the above will not be discussed here, as the first step, which follows from the Gaussian distribution above, is already rich and interesting.

Some years ago the problem of the statistics of contour lines in a random landscape [16,17] was studied by obtaining first the probability of a given trajectory, $\mathbf{r}(t)$ [18]. In a similar fashion the probability of a given trajectory is given by

$$\mathcal{P}\{\mathbf{r}(t)\} \propto \left\langle \prod_t \delta(\dot{\mathbf{r}}(t) - \mathbf{v}(\mathbf{r}(t),t)) \right\rangle, \tag{10}$$

where the average is over the distribution of the velocity field. Introducing the standard representation of the $\delta$ function and a vector function, $\mathbf{k}(t)$, we write

$$\mathcal{P}\{\mathbf{r}(t)\} \propto \int D\mathbf{k}(t) \left\langle \exp[i\int dt \mathbf{k}(t)\cdot(\dot{\mathbf{r}}(t) - \mathbf{v}(\mathbf{r}(t),t))] \right\rangle, \tag{11}$$

where $D$ denotes path integration. We **define** $\mathbf{k}(t)$ at the two end points $t=0$ and $t=T$ to be zero. Integration by parts replaces $\int dt \mathbf{k}(t) \cdot \dot{\mathbf{v}}(\mathbf{r}(t),t)$ by $\int dt \dot{\mathbf{k}}(t) \cdot \mathbf{v}(\mathbf{r}(t),t)$. The average is then performed to yield

$$\mathcal{P}\{\mathbf{r}(t)\} = \int D\mathbf{k}(t) \exp[-\frac{1}{6}\int dt_1 dt_2 \phi(|\mathbf{r}_{12}|,|t_{12}|)\dot{\mathbf{k}}(t_1)\cdot\dot{\mathbf{k}}(t_2) + i\int dt \mathbf{k}(t)\cdot\ddot{\mathbf{r}}(t)] \ . \tag{12}$$

The Fourier transform of the velocity correlation of an incompressible turbulent liquid, with an extreme infrared spatial noise, has been obtained by us a few years ago in the limit of long times and large wave length [14],

$$\langle \mathbf{v}(\mathbf{q},t)\cdot\mathbf{v}(\mathbf{p},0)\rangle = A\delta(\mathbf{q}+\mathbf{p})q^{-11/3}\exp(-Bqt^{3/2}). \tag{13}$$

This detailed form is in accord with Kolmogorov who gives in the inertial range

$$\langle[\mathbf{v}(\mathbf{r}_1,t_1)-\mathbf{v}(\mathbf{r}_2,t_2)]^2\rangle = A'|\mathbf{r}_{12}|^{2/3}f(B'|t_{12}|^{3/2}/|\mathbf{r}_{12}|), \tag{14}$$

where the scaling function $f(x)$ tends to a positive constant as $x$ tends to zero and $f(x) \propto x^{2/3}$ for large $x$. Using this inertial range form for the average in equation (15) above is justified as long as $|\mathbf{R}|\Lambda \ll 1$. The correlations we need are

$$\phi(|\mathbf{r}_{12}|,t_{12}) = \langle|\mathbf{v}|^2\rangle + \frac{1}{2}A'|\mathbf{r}_{12}|^{2/3}f(B't_{12}^{3/2}/|\mathbf{r}_{12}|), \tag{15}$$

where the first term on the left hand side is just the local velocity fluctuation (which is infinite for the extreme infra red spatial noise but is obviously finite in reality). The constant term in the correlation above does not contribute to the integral $\int dt_1 dt_2 \phi(|\mathbf{r}_{12}|,t_{12})\dot{\mathbf{k}}(t_1)\cdot\dot{\mathbf{k}}(t_2)$ appearing in equation (12) and can therefore be dropped). Therefore, the fact that in the limit of extreme infrared spatial noise, the correlation diverges, does not change the fact that the integral $\int dt_1 dt_2 \phi(|\mathbf{r}_{12}|,t_{12})\dot{\mathbf{k}}(t_1)\cdot\dot{\mathbf{k}}(t)$ exists even in that limit. Assume now that we are dealing with the case of a compressed trajectory, $\mathbf{R}^2/\langle R^2(T)\rangle \ll 1$. In this case the argument of $f$ appearing in equation (14) is large and consequently the probability density for the trajectory is given by

$$\mathcal{P}\{\mathbf{r}(t)\} = \int D\mathbf{k}(t)\exp[-\frac{C}{2}\int dt_1 dt_2 |t_{12}|\dot{\mathbf{k}}(t_1)\cdot\dot{\mathbf{k}}(t_2) + i\int dt \mathbf{k}(t)\cdot\ddot{\mathbf{r}}(t)] \ . \tag{16}$$

Integrating by parts the double integral, find that it equals $C\int dt \mathbf{k}^2(t)$. Now the path integral can be performed easily to yield equation (7) for $\mathcal{P}$ and the resulting equation (8) for the trajectory end point distribution with $Q$ replaced by $2C$.

When $\mathbf{R}^2/\langle R^2(T)\rangle \gg 1$, the trajectories are stretched and the situation is more interesting. We will present here a crude approximation, which yields the behavior at the tail of the trajectory end point distribution. We have obtained a more refined but also more complicated derivation, which will not be presented here because of space limitations. Both derivations yield essentially the same result. It is safe to assume that if a trajectory is stretched most of its segments are also stretched so that $f$ in equation (15) can be replaced by a constant. Clearly each segment of a stretched trajectory is stretched differently. We will ignore that difference and assume that all segments are stretched in the same way, which is therefore the stretching of the whole trajectory. Using this approximation we write

$$|\mathbf{r}_{12}|^{2/3} = [|t_{12}|/T]|\mathbf{R}|^{2/3} \ . \tag{17}$$

Consequently $C$ in equation (16) is replaced by $\xi = c|\mathbf{R}|^{2/3}/T$. Thus a direct consequence of equation (8) is the tail behavior

$$P(\mathbf{R},T) \propto \exp[-\mathbf{R}^2/4\xi T^3] = \exp[-(|\mathbf{R}|/T^{3/2})^{4/3}/4c] \ . \tag{18}$$

Note that the only difference between the compressed and the stretched situation arises due to the difference in the behavior of $f(x)$ for small and large values of $x$. In fact, within the same approximation, we could interpolate, assuming we know the form of the scaling function, $f$. What we have to do is to use equation (12) and replace the argument of $f$ in equation (14) by $B'T^{3/2}/|\mathbf{R}|$. Consequently the interpolated expression reads

$$P(\mathbf{R},T) \propto \exp[-3f(B'T^{3/2}/|\mathbf{R}|)(|\mathbf{R}|/T^{3/2})^{4/3}/4A']. \tag{19}$$

While the compressed case could be viewed as arising from random impulses, the strength of those impulses is modified and becomes weaker as the trajectory become more stretched.